\documentclass[11pt,twoside]{article}

\usepackage{asp2006}
\usepackage{epsf}
\usepackage{psfig}
\usepackage{lscape}
\usepackage{graphicx}
\font\piedi=cmr8

\def\spb{\smallskip\par\noindent $\bullet\;$}
\catcode`\@=11
\def\gsim{\ifmmode{\mathrel{\mathpalette\@versim>}}
    \else{$\mathrel{\mathpalette\@versim>}$}\fi}
\def\lsim{\ifmmode{\mathrel{\mathpalette\@versim<}}
    \else{$\mathrel{\mathpalette\@versim<}$}\fi}
\def\@versim#1#2{\lower 2.9truept \vbox{\baselineskip 0pt \lineskip 
    0.5truept \ialign{$\m@th#1\hfil##\hfil$\crcr#2\crcr\sim\crcr}}}
\catcode`\@=12
\def\msun{\hbox{$M_\odot$}}

\begin{document}
\title{Very Massive Galaxies: A Challenge For Hierarchical Models?}   
\author{Alvio Renzini}   
\affil{INAF -- Osservatorio Astronomico di Padova, Italy}    

\begin{abstract} 
Hierarchical models of galaxy formation now provide a much closer
match to observations than they did a few years ago. The progress has
been achieved by adjusting the description of baryonic processes such
as star formation and supernova/AGN feedback, while leaving the
evolution of the underlying dark matter (DM) halos the same.  Being
most results very sensitive to the input baryonic physics, the
ultimate vindication of the hierarchical paradigm should come from
observational tests probing more directly the merging history of DM
halos rather than the history of star formation. Two questions may
start addressing this deeper level: is the predicted halo merging rate
consistent with the observed galaxy merging rate? and, are predicted
and observed evolution of the galaxy mass function consistent with
each other? The current status of these issues is briefly reviewed.

\end{abstract}


\section{Introduction}

The organizers of this meeting have asked me to review the current
status of the intercomparison of observations and theory of galaxy
formation and evolution, specifically for the massive galaxies.  This
is not an easy task, as both observations and theory are progressing
at a virtually daily rate, and what is perceived as a major discrepancy
today may soon disappear.  Indeed, pressed by major advancements in mapping
galaxy populations at low and high redshifts, theoretical models of
galaxy formation have evolved dramatically in the last few years. By
and large, models are now in far better agreement with observations
that were just a few years ago, especially concerning the evolution of
massive, passively evolving galaxies. Until recently most models
failed to produce such {\it early type galaxies} (ETG) in the observed
number virtually at all redshifts, primarily because star formation
was continuously fed by cooling flows in massive dark-matter halos
(e.g. Somerville 2004).  As a result, models did not reproduce the
color bimodality in galaxy color-magnitude diagrams so evident at
$z=0$ (e.g., Baldry et al. 2004), as well as all the way to $z\sim 1$,
as illustrated here in Fig. 1 (from Scarlata et al. 2007).

On the other hand, it was understood since a long time that cooling
flows could not be ubiquitous among local ETGs, otherwise such
galaxies would have been ten or more times brighter in X-Rays than
they typically are (Ciotti et al. 1991). Suppression of stationary
cooling flows was then achieved by invoking feedback from a central
supermassive black hole, and physical models of such AGB feedback were
attempted (Ciotti \& Ostriker 1997, 2001). Moreover, evidence for the
bulk of stars in ETGs having formed at very high redshifts was
available since at least the early 'nineties (cf.  Renzini 2006, for a
recent review). To match these constraints, models have been tuned to
promote more star formation at early cosmic times, and to suppress it
in massive halos not too much after the bulk of stars had formed.

The first goal was achieved by adding a {\it starburst} mode of star
formation to the {\it quiescent} one, i.e., by assuming that a
fraction of the gas in merging galaxies is instantly turned into stars
(e.g., Kauffmann \& Haehnelt 2000; Somerville, Primack, \& Faber
2001). For the second goal, supernova feedback was adjusted to higher
values, and when not enough star formation was suppressed altogether
attributing such effect to AGN feedback, as first incorporated into
semi-analytical models by Granato et al. (2001). AGN feedback is now
included in most galaxy formation models, although one should be aware
that no physical modeling of the process is actually attempted. In
practice, star formation is discontinued when appropriate to reproduce
pertinent observations, and the suppression is attributed to AGN
feedback.

\begin{figure}[!t]
\begin{center}
\includegraphics[width = 8cm, angle=-90]{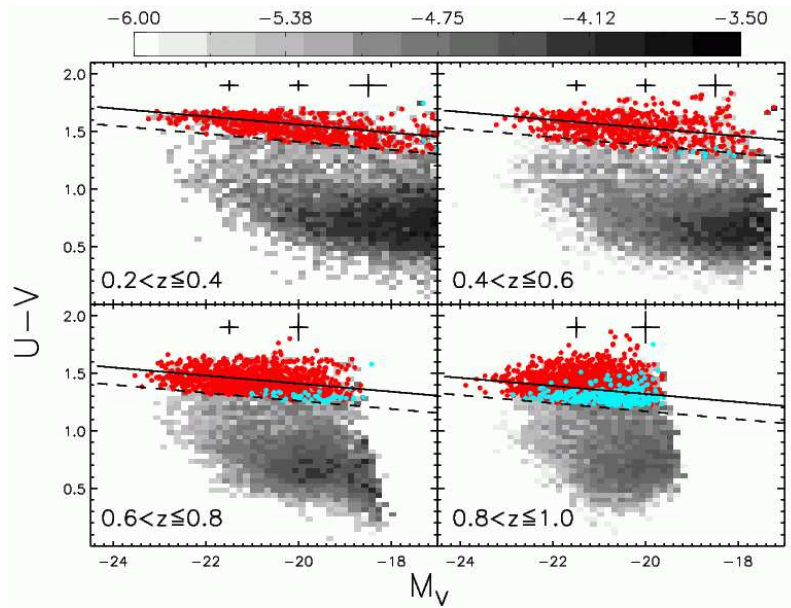}
\caption{\piedi The rest-frame $(U-V$ vs $M_{\rm V}$ color-magnitude
diagram for galaxies in the COSMOS field, over four redshift bins. The
band bounded by the dashed lines includes the red sequence of
early-type galaxies. Red points refer to galaxies whose SED is
best-fitted by a purely passive template, while cyan points refer to
objects whose SED indicates still ongoing star formation.  From
Scarlata et al. (2007)}
\end{center}
\end{figure}
 
As a result of these improvements, models are now in far better
agreement with observations than they used to be. However, the lesson
to be drawn from this progress is that the physics of the baryonic
component of the universe is so complex, that the theoretical
description of its manifold behaviors (star formation, supernova
feedback, black hole formation and growth, AGN feedback, etc.)
necessarily relies on schematic recipes and numerous adjustable
parameters. Thus, empirical star formation histories don't really give
us a test that models could not be forced to match.  Therefore, the
question is whether there is a more fundamental test, an ultimate
vindication for the hierarchical CDM paradigm of galaxy formation.

Contrary to baryonic physics, the growth of dark matter halos from the
initial perturbations is governed by much simpler physics, and the
merging history of such halos can be reliably predicted by modern
N-body simulations (e.g., Springel, Frenk \& White 2006).  Hence, the
{\it Mass Assembly} of galaxies may provide the more fundamental test of the
CDM paradigm that we are looking for, as it is more directly related
to the merging history of DM halos.

\section{The rate of (dry)  merging among massive galaxies}  

Major merging events, and especially ETG-ETG {\it dry} mergers at
$z<1$, have been proposed to be responsible for the mass assembly of
ETGs to their final size (e.g., Bell et al. 2004: Faber et al. 2005).
Therefore, the direct estimate of the rate of (dry) merging events is
potentially able to check this suggestion, and perhaps even to provide
a test for the CDM paradigm itself.  So, do empirical merging rates
among the most massive galaxies support this suggestion, and how do
they compare with models?

There have been several recent attempts at estimating this rate, both in
the nearby universe and all the way to redshift $\sim 1$. The main
conclusion of these studies is synthetically reported here:

\spb ``less than 9\% of massive ETGs experienced a major merger since
     $z=1.2$'' \par
     (DEEP2, Lin et al. 2004) 
\spb ``$\sim 35\%$ of ETGs
     had a merger with a $>1:4$ mass ratio since $z=0.1$''\par
     (van Dokkum 2005) 
\spb ``Each ETGs has undergone 0.5-1 major dry mergers since $z=0.7$''\par
     (COMBO-17, Bell et al. 2006a) 
\spb ``$\sim 20\% $ of $M_*>2.5\times 10^{10}\msun$ had a major dry
     merger since $z=0.8$ "\par
     (COMBO-17, Bell et al. 2006b) 
\spb ``less than 1\% probability of a dry merger per Gyr at $z<0.36$''\par
     (SDSS, Masjedi et al. 2006).

While the different merger definitions and redshift intervals
complicate the intercomparison, these derived merging rates differ too
much for all being consistent with one another. Arguing
whether some of these estimates are more reliable than others goes
beyond the scope of this review. Suffice here to point out that
major discrepancies exist among the most recent estimates of the
merging rate, hence they do not provide yet a clear cut test for
the CDM paradigm.

\section{The Galaxy Mass Function From $z\sim 1$ to $z=0$}    

An alternative way to check for the mass assembly of galaxies is to
look to the evolution with redshift of their mass function, or of its
proxy, the luminosity function. Within the hierarchical paradigm the
most massive galaxies are the last to be fully assembled, and their
number is predicted to keep increasing with time all the way to the
present. We all certainly agree that the most massive galaxies must
have originated from the highest primordial overdensities, hence
having been the first to start forming stars. If a mechanism exists
such that the more massive a galaxy the sooner star formation is
quenched, then the more massive a galaxy the older its stellar
content, and {\it downsizing} follows ``naturally'', as indicated by
the observations (again, cf. Renzini 2006, for an extensive review.)
However, along with downsizing in star formation, the continuous
merging of dark matter halos should result in an {\it upsizing} in
mass assembly, with the most massive galaxies being the last to be
fully assembled.

\begin{figure}[!t]
\begin{center}
\includegraphics[width = 13cm]{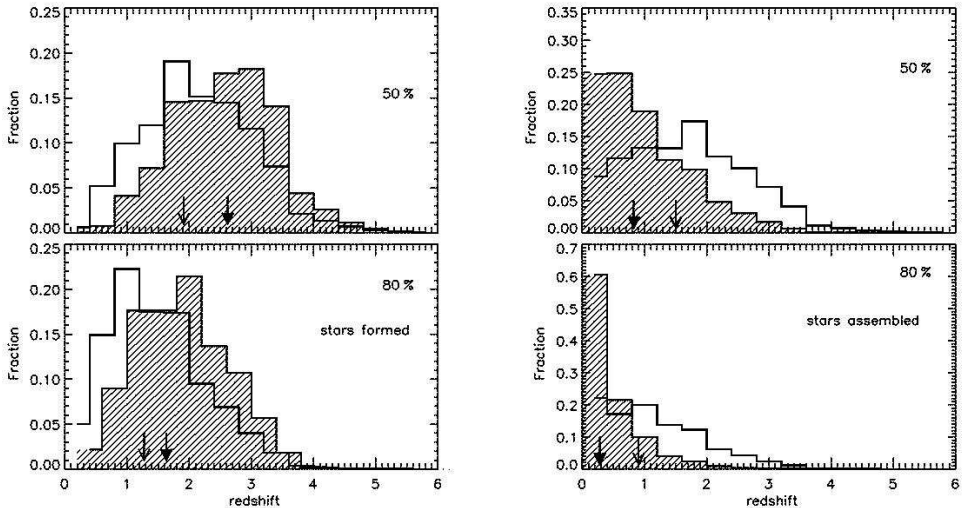}
\caption{\piedi Left panel: The distribution of the formation
redshift for stars in model early type galaxies (De Lucia et
al. 2006). Histograms give the fraction of model galaxies at $z=0$
that have completed 50\% (upper panel) or 80\% (lower panel) of their
star formation at the various redshifts. The open and shaded
histograms refer to galaxies with stellar mass at $z=0$ in excess of
$4\times 10^9\msun$ and $10^{11}\msun$, respectively. Right Panel: The
same, but for the stellar mass assembly in one object. The most
massive galaxies generally form their stars earlier, but assemble them
later via merging, compared to less massive galaxies.}
\end{center}
\end{figure}

These trends are illustrated by the result of recent semi-analytical
models shown in Fig. 2 (from De Lucia et al. 2006), where indeed
downsizing in star formation and upsizing in mass assembly are clearly
apparent. Now, while there is direct evidence for downsizing in star
formation, at low as well as high redshift, do we have observational
evidence also for upsizing in mass assembly? Here the situation is
less clear cut.  Fig. 3 (From Cimatti et al. 2006) shows that the
number density of ETGs with stellar mass in excess of $\sim
10^{11}\msun$ stays nearly constant up to $z\sim 0.7$ and starts
decreasing only at higher redshifts. On the contrary, the number
density of less massive ETGs declines steadily with redshift, from
$z=0$ onward. Cimatti et al. argued that in the hierarchical
models shown in Fig. 2 the most massive ETGs are the most rapidly
declining component, contrary to empirical evidence. Shall we conclude
that Fig. 3 provides evidence against the upsizing predicted by the
hierarchical merging paradigm? Such a conclusion may be
premature. Indeed, also shown in Fig. 3 are the predictions from
another rendition of the paradigm (Menci et al. 2006, supplemented by
private communication), where the number density of the most massive
ETGs does stay constant all the way to $z\sim 0.8$, in nice agreement
with the observations.  The baryonic recipes differ somewhat between
the two sets of models, and it remains to be understood which specific
ingredient is responsible for the different predictions.

\begin{figure}[!t]
\begin{center}
\includegraphics[width = 8cm]{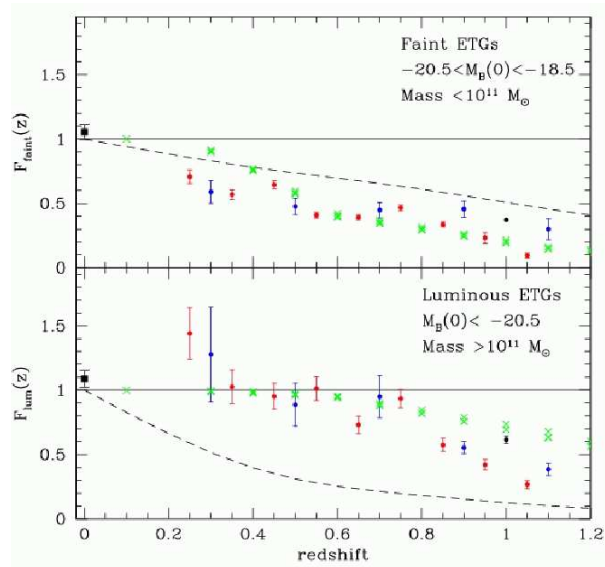}
\caption{\piedi (Left) The evolution of the number density of ETGs
normalized to their density at $z=0$ from the COMBO-17 (Bell et
al. 2004) and DEEP2 (Faber et al. 2005) surveys, as from Cimatti et
al. (2006). The upper and lower panel refer to galaxies whose stellar
mass is below or above $\sim 10^{11}\msun$, respectively. The dashed
lines refer to the fraction of model ETGs from De Lucia et al. (2006)
which have assembled 80\% of their stellar mass, as inferred from
Fig. 2. Crosses refer to the models of Menci et al. (2006).}
\end{center}
\end{figure}

\begin{figure}[!t]
\begin{center}
\includegraphics[width = 8cm]{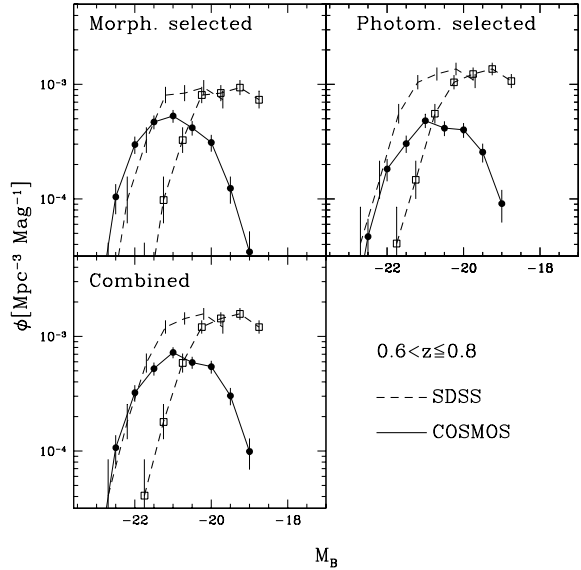}
\caption{\piedi (Right) The rest-frame $B$-band luminosity function
for COSMOS samples of ETGs at $z=0.7$ selected either morphologically
or photometrically, or combining the two criteria are shown as filled
circles (from Scarlata et al. (2007). The dashed line with open
squares refer to the local luminosity function from SDSS, which is
also shown solidly brightened by 0.95 magnitudes so to mimic the
expected effect of passive evolution. Note that there is virtually no
number density evolution at the top end of the luminosity function,
whereas the number density of faint ETGs drops dramatically at high
redshift.}
\end{center}
\end{figure}

One may argue that limiting the test to ETGs  may introduce a
bias, as in some simulations the massive galaxies may
be fully assembled at $z\sim 0.7$, but their star formation has not
been quenched in time to qualify them as ETGs. On the observational
side, even very modest amounts of ongoing star formation can suffice
to disqualify a galaxy from listing in morphological or photometric
samples of ETG. On the theoretical side, no strong prediction can be
made about modest residual star formation within massive DM halos.
Note, however, that virtually all most massive galaxies ($M_*>$ few
$10^{11}\msun$) are passive ETGs all the way to $z\sim 1$, as apparent
from Fig. 1, and their number density has not appreciably changed
since $z\sim 1$ (e.g. Bundy et al. 2006; Borch et al. 2006).  Thus,
Cimatti et al. also argued that if the number density of the most
massive ETGs is constant up to $z\sim 0.8$, then very little room is
left for dry merging playing a major role in the build-up of massive
ETGs in the corresponding interval of cosmic time. A similar
conclusion is indeed reached by Scarlata et al. (2007) for galaxies in
the COSMOS field, as illustrated here in Fig. 4.

In any event, the total stellar mass of a galaxy is somewhat more
fundamental than either its morphology or color, and should be a
more robust prediction of the models. Hence, looking at the evolution
of the mass function may offer a chance to submit the models to a
deeper scrutiny. Still, the mass functions from different renditions
of the CDM paradigm, once tuned to match low redshift observables used
to diverge dramatically beyond redshift $\sim 1$ (e.g., Fontana et
al. 2004), due just to different assumption concerning the {\it
baryonic} physics.

\begin{figure}[!t]
\begin{center}
\includegraphics[width = 6cm, angle = -90]{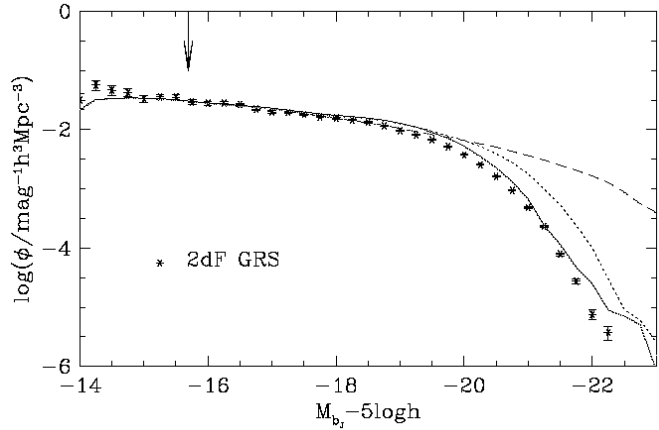}
\caption{\piedi The luminosity function of model galaxies at $z=0$
from Bower al. (2006). The dashed and dotted lines represent the case
when AGN feedback is switched off and on, respectively; the continuous
line corresponds to the further assumption of internal obscuration by
dust. The data point are from the SDSS. }
\end{center}
\end{figure}

In a recent attempts at tracing the evolution of the mass function of
galaxies all the way to $z\sim 4$, broad agreement with the results of
several semi-analytic as well as hydrodynamical models was instead
found (Fontana et al. 2006), which illustrates the great progress on
the theoretical side that has been achieved in the last two years. Yet, not
all problems may have been solved.

Here I will focus on three recent theoretical contributions,
where subtle differences with respect to observations may guide us
towards the fundamental test we are looking for. Earlier generations
of models suffered from an extreme overproduction of massive galaxies
at $z=0$ coupled with an extremely rapid disappearance of them with
increasing redshift (e.g. Baugh et al. 2003, see their Fig. 4), at
dramatic variance with observations. The progress made in the meantime
is here illustrated in Fig. 5, from Bower et al. (2006), and
resulting from an update of the same semi-analytic code used by Baugh
et al. (2003).  Suppressing star formation in massive halos,
attributed to AGN feedback, has a dramatic effect at the top end of
the luminosity function (LF), thus cutting the previous 
overproduction of massive galaxies by $z=0$, while the same model gives a
reasonable fit to the observed LF up to $z\sim 1.5$.  Still, a
sizable excess is noticeable at the top end of the $b_{\rm
J}$-band LF at $z=0$, which is finally adjusted by assuming internal dust
obscuration in the model galaxies. As well know, the top end of the LF
is dominated by passively-evolving galaxies all the way to $z\sim 1$
(cf. Fig. 1), for which internal extinction must be very low. Hence,
this model may still somewhat overproduce massive galaxies at low redshifts.

A relatively modest overproduction of massive galaxies at $z=0$ is
also apparent in a recent incarnation of the Millennium Simulation
(Kitzbichler \& White 2006), here shown in Fig. 6. Besides, the model
tend to underproduce massive galaxies at higher and higher redshifts,
compared to the observed mass functions taken at face
value. Observations are then brought to agreement with the models by
convolving the theoretical mass function with an assumed random error
in empirical mass determinations. Given the very steep slope at high
masses, random errors have a fairly appreciable  effect at the top end,
while having very little effect further down in the mass function.
Thus, the discrepancy is alleviated by attributing it to an empirical
overestimate of the number of massive galaxies at high redshifts.
However, applying a similar convolution also at $z=0$ would further
amplify the overproduction of the most massive galaxies in the local
universe.
\begin{figure}[!b]
\begin{center}
\includegraphics[width = 8cm]{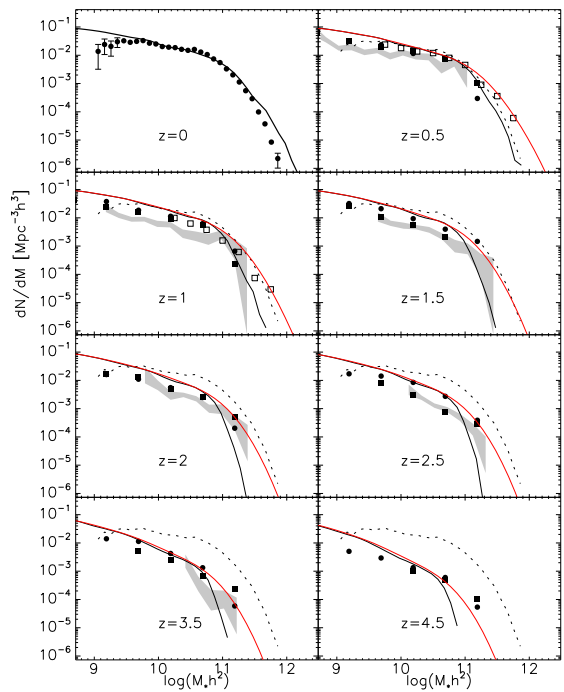}
\caption{\piedi (Left) The evolution of the stellar mass function of model
galaxies from Kitzbichler \& White (2006). Data at z=0 are from the
SDSS, then identically repeated as a dashed line in the other
panels. Higher redshift data are from Drory et al. (2005, symbols) and
from Fontana et al. (2006, shaded area). The lower and upper solid
lines represent the model result before and after convolution of the
mass function with an assumed random error 0.25 dex in empirical mass
determinations. No such correction was applied at $z=0$.}
\end{center}
\end{figure}

\begin{figure}[!b]
\begin{center}
\includegraphics[width = 8cm]{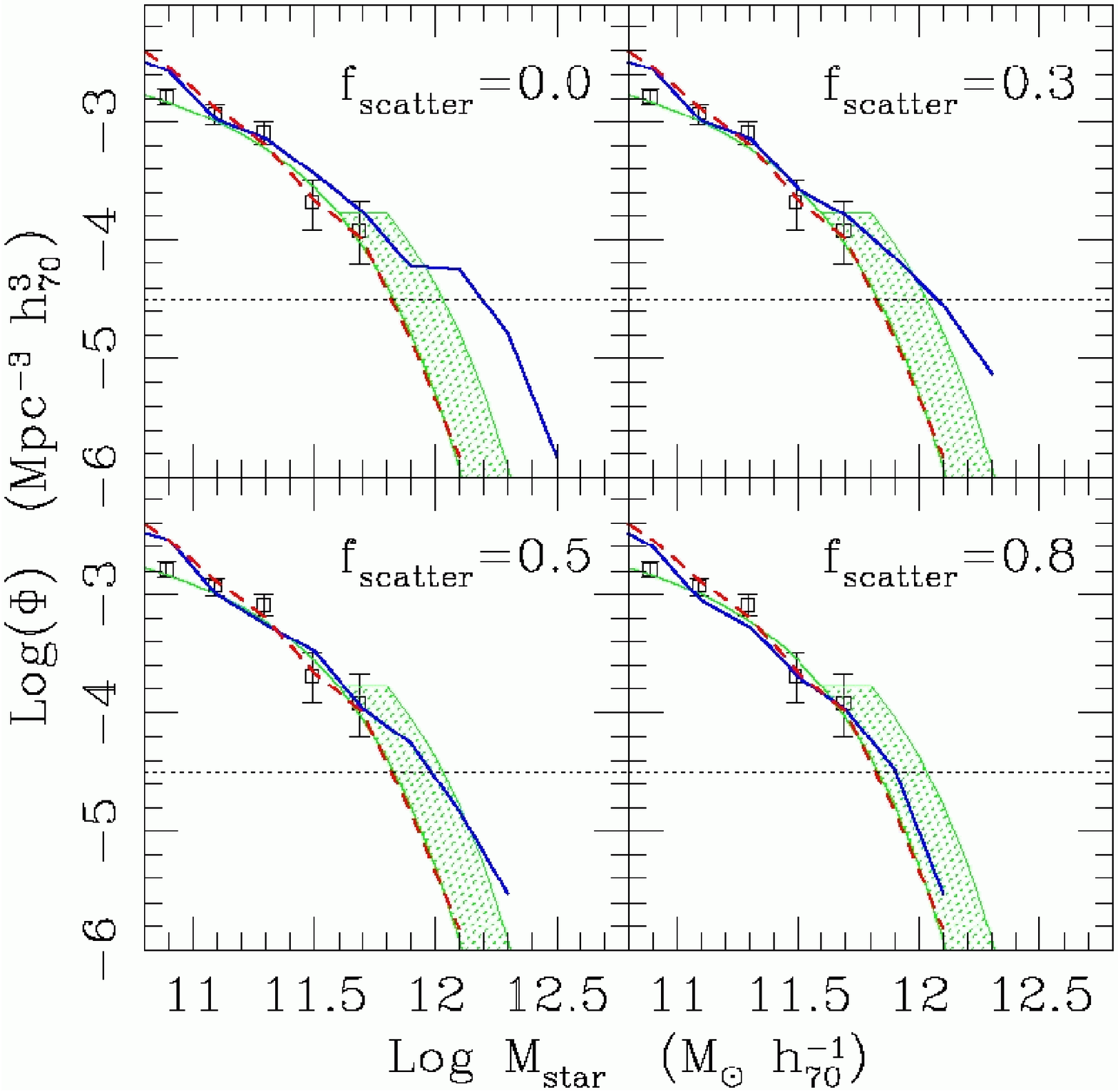}
\caption{\piedi (Right) The evolution from $z=1$ (dashed line) to
$z=0$ (solid line) of the stellar mass function of model galaxies from
Monaco et al. (2006). In the models star formation was completely
switched off from $z=1$ to 0.  In each panel $f_{\rm scatter}$ is the
assumed fraction of the stellar mass of the merging galaxy which is
assumed to be lost in the intergalactic medium. Data points are from
Fontana et al. (2006). The upper envelope of the shaded area
corresponds to the observed mass function at $z=0$.}
\end{center}
\end{figure}

How to cope with the model overproduction of massive galaxies between
$z=1$ and $z=0$ is the prime focus of the third study (Monaco et
al. 2006), with Fig. 7 illustrating the case. Having fixed the star
formation and feedback parameters in such a way to get a good fit to
the mass function at $z=1$, Monaco et al. continued the simulation
down to $z=0$ having switched off all star formation, and allowing the
mass of individual galaxies to grow under the sole effect of
merging. By ignoring any possible mass increase by star formation, the
result should be a {\it lower} limit to the top end of the mass
function, and yet by $z=0$ an overproduction at the top mass end by
more than an order of magnitude has gown (cf. top/left panel in
Fig. 7). Monaco et al. explore the possibility of solving the
discrepancy by assuming that a fraction $f_{\rm scatter}$ of the
stellar mass of merging galaxies does not end up in the merger
product, but is rather scattered away to make up a population of
free-floating intergalactic stars. Monaco et al.  conclude that with
$f_{\rm scatter}$ at least as high as $\sim 0.5$ the theoretical mass
function at $z=0$ could be reconciled with the observed one. One may
also argue from Fig. 7 that a value as high as $f_{\rm scatter}=0.8$
would give an even better fit to the data, which would be almost
equivalent to neglect merging altogether.

In summary, 2006 semi-analytic models still appear to overproduce the
most massive galaxies at $z=0$, although different recipes have been
suggested to bring them in agreement with the observations, namely,
internal obscuration, random errors in empirical estimates of the
masses of galaxies, and a relative inefficiency of merging in growing
galaxy masses. In essence, there appears to be less merging activity
in nature between $z=1$ and $z=0$, compared to CDM expectations, which
may also be indicated by some of the attempts at directly estimate the
merging rates (cf. Section 2). May be there is a fourth option to
solve the apparent discrepancy: perhaps dark matter halos {\it do
merge} after all, but most galaxies don't, and just the {\it halo
occupation number} increases.

\section{Conclusions}

The last generation of hierarchical CDM models of galaxy formation and
evolution provide a far better match to observations compared to the
recent past. This improvement has resulted mainly from tuning the star
formation and feedback recipes which are used to treat  the {\it
baryonic physics} in a schematic way.

A more fundamental test of the CDM paradigm should come from direct estimates
of the merging rate of massive galaxies, to be compared to the
theoretical merging rate of DM halos. However, current estimates of
the (dry) merging rate at $z<1$ diverge widely, hence are not yet able
to give a secure answer.

There appears to be some evidence for downsizing in both star
formation and mass assembly of ETGs, i.e. the most massive ellipticals
seem to be the first, not the last to be fully assembled as instead
expected on the theoretical side.  Once tuned to match the LF/MF at
$z\sim 1$, models apparently overproduce the most massive galaxies
($M_*>$ few $10^{11}\msun$) by z=0, due to still high merging rate of
DM halos between $z\sim 1$ and $z=0$. It has been suggested that the
discrepancy may not be real, but just arising from observational
errors in stellar mass estimates. In alternative, if the discrepancy
is real it could be due to merging not automatically resulting in an
additive increase of the stellar mass of galaxies. Or maybe the
merging rate of galaxies differs from the merging rate of their host
halos.

With so many degrees of freedom in our description of baryonic
physics, the vindication of the CDM hierarchical paradigm of galaxy formation
should rather come from 
accurate estimates of the  evolution with redshift of the galaxy merging rate, 
along with the evolution of the mass function, and especially of its top end.
\noindent
\acknowledgements I am grateful to the organizers for having forced me
to look deeper at the comparison of galaxy formation models and
observations. I am indebted to the authors of the papers from which
figures have been reproduced here with their permission.



\begin{thebibliography}{}
\bibitem[]{} Baugh, C.M., Benson, A.J.. Cole, S., Frenk, C.S., Lacey, C. 2003,
             in The Mass of galaxies at Low and High Redshift, ed. R. Bender 
              \&  A. Renzini (Berlin: Springer), p. 91
\bibitem[]{} Bell, E.F., Wolf, C., Meisenheimer, K., Rix, H.-W., et 
           al. 2004, ApJ 608, 752
\bibitem[]{} Bell, E.F., Naab, T., McIntosh, D.H., Somerville, R.S.,  
             et al. 2006a, ApJ 640, 241
\bibitem[]{} Bell, E.F., Phleps, S., Somerville, R.S., Wolf, C., Borgh, A., et 
             al. 2006b, ApJ, in press (astro-ph/0602038)
\bibitem[]{} Borch, A., Meisenheimer, K., Bell, E.F., Rix, H.-W., et
             al. 2006, A\&A, 453, 869 
\bibitem[]{} Bower, R.G., Benson, A.J., Malbon, R., Helly, J.C., 
             et al. 2006, MNRAS, 370, 654
\bibitem[]{} Bundy, K., Ellis, R.S., Conselice, C.J., Taylor, J.E., 
             et al. 2006. ApJ, 651, 120
\bibitem[]{} Cimatti, A., Daddi, E., \& Renzini, A. 2006, A\&A, 453, L29
\bibitem[]{} Ciotti, L., D'Ercole, A., Pellegrini, S., Renzini, A. 1991,
             ApJ, 376, 380
\bibitem[]{} Ciotti, L., \& Ostriker, J.P. 1997, ApJ, 487, L10
\bibitem[]{} Ciotti, L., \& Ostriker, J.P. 2001, ApJ, 551, 131
\bibitem[]{} De Lucia, G., Springel, V., White, S.D.M., Croton, D., \& 
             Kauffmann, G. 2006. MNRAS 366, 499
\bibitem[]{} Drory, N., Salvato, M., Gabasch, A., Bender, R., Hopp, U., et al. 
             2005, ApJ, 619, L131
\bibitem[]{} Faber, S.M., Willmer, C.N.A., Wolf, C., Koo, D.C.,  
             et al. 2006, astro-ph/0506044
\bibitem[]{} Fontana, A., Pozzetti, L., Donnarumma, I., Renzini, A., 
             et al. 2004, A\&A, 424, 23
\bibitem[]{} Fontana, A., Salimbeni, S., Grazian, A., Giallongo, E., 
             Pentericci, L., et al. 2006, astro-ph/0609068
\bibitem[]{} Granato, G.L., Silva, L., Monaco, P., Panuzzo, P.,  
             et al. 2001, MNRAS, 324, 757
\bibitem[]{} Kauffmann, G., Haehnelt, M. 2000, MNRAS, 311. 576 
\bibitem[]{} Kitzbichler, M.G., \& White, S.D.M. 2006, MNRAS, 366, 858
\bibitem[]{} Lin, L., Koo, D.C., Willmer, C.N.A., Patton, D.R.,  
             et al. 2004, ApJ, 617, L9
\bibitem[]{} Masjedi, M., Hogg, D.W., Cool, R.J., Eisenstein, D.J.,
             et al. 2006, ApJ, 644, 54
\bibitem[]{} Menci, N., Fontana, A., Giallongo, E., Grazian, A., \&
             Salimbeni, S. 2006, ApJ, 647, 753
\bibitem[]{} Monaco, P., Murante, G., Borgani, S., \& Fontanot, F. 2006, 
             ApJ, 652, L89
\bibitem[]{} Renzini, A. 2006, ARA\&A, 44, 141
\bibitem[]{} Scarlata, C., Carollo, C.M., Lilly, S.J., Feldmann, R., Kampczyk, 
             P., et al. 2007, ApJS, in press (astro-ph/0701746)
\bibitem[]{} Somerville, R.S. 2004, in Multiwavelength Mapping of
             Galaxy Formation and Evolution, ed. A. Renzini \& R. Bender 
             (Berlin: Springer), p. 131
\bibitem[]{} Somerville, R.S., Primack, J.R., \& Faber, S.M. 2001, MNRAS, 
             320, 504
\bibitem[]{} Springel, V., Frenk, C.S., \& White, S.D.M. 2006, Nature,
             440, 1137
\bibitem[]{} van Dokkum, P.G. 2005, AJ, 130, 2647

\end{thebibliography}
\end{document}